\documentclass[12pt,a4paper]{article}
\usepackage[utf8]{inputenc}
\usepackage{graphicx,hyperref}
\usepackage{amsmath,amssymb,amsfonts}
\usepackage[a4paper,left=2.5cm,right=2.5cm,top=2.5cm,bottom=2.5cm]{geometry}
\usepackage[sort&compress,numbers]{natbib}
\usepackage[affil-it]{authblk}
\usepackage{xcolor}

\def\Lag{{\mathcal L}{}}

\def\Mag{{\mathcal M}{}}

\def\Lw{{\stackrel{\bullet}{{\mathcal L}}}{}}
\def\Gammaw{{\stackrel{\bullet }{\Gamma}}{}}
\def\Gammawz{{}^0{\stackrel{\bullet }{\Gamma}}{}}
\def\Tw{{\stackrel{\bullet}{T}}{}}
\def\Twz{{}^0{\stackrel{\bullet }{T}}{}}
\def\omw{{\stackrel{\bullet}{\omega}}{}}
\def\omwz{{}^0{\stackrel{\bullet }{\omega}}{}}
\def\Kw{{\stackrel{\bullet}{K}}{}}
\def\Kwz{{}^0{\stackrel{\bullet }{K}}{}}
\def\nablaw{{\stackrel{\bullet}{\nabla}}{}}

\def\ombol{{\stackrel{\circ}{\omega}}{}}
\def\Gammabol{{\stackrel{\circ}{\Gamma}}{}}

\def\nabbol{{\stackrel{\circ}{\nabla}}{}}

\def\Gammas{{\stackrel{\rule{3pt}{3pt}}{\Gamma}}{}}
\def\omw{{\stackrel{\bullet}{\omega}}{}}
\def\Qs{{\stackrel{\rule{3pt}{3pt}}{Q}}{}}
\def\Ls{{\stackrel{\rule{3pt}{3pt}}{L}}{}}
\def\Qsz{{}^0{\stackrel{\rule{3pt}{3pt}}{Q}}{}}

\def\SCO{{\mathbf{\Omega}}{}}
\def\LCO{{\mathbf{G}}{}}

\begin{document}
\title{\bf Teleparallel Gravity, Covariance and Their  Geometrical Meaning}
\author{Martin Kr\v{s}\v{s}\'ak\thanks{Electronic address: \texttt{martin.krssak@gmail.com, martin.krssak@fmph.uniba.sk}}}
\affil{Department of Theoretical Physics, Faculty of Mathematics, Physics and Informatics, Comenius University, Bratislava, 84248, Slovak Republic}
\affil{Department of Astronomy, School of Astronomy and Space Science, University of
	Science and Technology of China, Hefei, Anhui 230026, China}
\maketitle
\begin{abstract}
We explore the geometrical meaning of teleparallel geometries and the role of covariance in their definition. We argue that   pure gauge connections are a necessary ingredient for describing geometry and gravity in terms of  torsion and non-metricity. We show the other viable alternative is using the Einstein and M\o ller Lagrangians, but these are defined through the Riemannian connection coefficients and hence do not involve torsion nor non-metricity. We argue that the teleparallel geometries can be defined on the manifold without introducing  any additional structures and that they naturally provide the covariant framework for the Einstein and M\o ller Lagrangians. We explore some consequences of this viewpoint for the modified theories of gravity as well. 
\end{abstract}

\section{Introduction}
Teleparallel theories of gravity were originally proposed by Einstein in the late 1920s as an attempt for the unified field theory of gravity and electromagnetism \cite{Einstein1928a,Einstein1928b,Einstein1929a}. Since  1960s the idea of teleparallelism was revived as an alternative approach to gravity, which found many applications addressing various issues in gravity and cosmology. Among these were the improvements in definitions of gravitational energy-momentum \cite{Moller1961,Pellegrini1963,Moller1966,Moller1978}, formulation of general relativity as a  gauge theory \cite{Hayashi:1967se,Cho:1975dh,Hayashi:1977jd}, and more recently a plethora of modified gravity theories
with the aim to address dark energy and other problems in cosmology \cite{Ferraro:2006jd,Ferraro:2011ks,Linder:2010py,Geng:2011aj,Cai:2015emx,Bahamonde:2021gfp,Krssak:2018ywd}.

Nevertheless, we face a rather curious situation that  even after almost 100 years since the initial  formulation,  debates persist regarding the very foundations of teleparallel theories of gravity, which  primarily revolves around  the pure gauge nature of teleparallel connections, due to which these  connections can be transformed to zero and hence effectively eliminated from the theory.   The main question then concerns whether to consider these  pure gauge connections as a fundamental variable of the theory. 

If the non-trivial teleparallel connection is utilized,  teleparallel theories are covariant  and the elimination of the  connection is viewed only as a choice of a specific gauge \cite{Obukhov:2002tm,Lucas:2009nq,AP,Krssak:2015rqa,Krssak:2015oua,Krssak:2018ywd}. The other viewpoint is to consider the theory without the teleparallel connection as the fundamental one. Then teleparallel theories are viewed as fundamentally non-covariant \cite{Maluf:1994ji,Maluf:2013gaa,Maluf:2018coz}, and  the covariant formulation is considered only as an artificial restoration of covariance using the  St\"uckelberg trick \cite{BeltranJimenez:2017tkd,BeltranJimenez:2018vdo,BeltranJimenez:2019nns,BeltranJimenez:2019esp,Golovnev:2023yla}. See \cite{Li:2010cg,Golovnev:2017dox,Bejarano:2019fii,Blagojevic:2023fys} for further discussion in the modified case.

Our goal in this paper is to address the origins of the issue of covariance in teleparallel theories of gravity. We first review Einstein's original motivation and his definition of teleparallelism, as well as how teleparallelism was revived starting in the 1960s. We also explain the relationship of these theories to the so-called Einstein and M\o ller Lagrangians, both of which play important roles in our discussion. Then, we show that there exists a generalization of the original notion of teleparallelism, where teleparallel geometries are viewed as special limiting cases of metric-affine geometries, naturally introducing the pure gauge connections. These two different notions of teleparallelism are then compared, and we demonstrate that when we choose to gauge away the pure gauge connection, we obtain the original Einstein's teleparallelism within the metric affine approach.

Our main argument in this paper is that both approaches are equally suitable for performing calculations and constructing gravitational theories. However, the pure gauge connections are crucial elements to attribute non-trivial geometry to torsion or non-metricity, as without them, torsion and non-metricity are not true tensors. On the other hand, if we view these gauge-fixed geometries as the fundamental definitions of teleparallel theories, we are actually  using the coefficients of anholonomy and partial derivatives of the metric to describe the geometry and gravity.

We suggest that this does not actually represent a problem, as long as we do not misidentify these objects as torsion or non-metricity tensors. We argue that describing gravity through the coefficients of anholonomy or partial derivatives of the metric is fully consistent and equivalent to using the Einstein and M\o ller Lagrangians, and we argue in their favor. However, these Lagrangians are formulated within the Riemannian geometry using the Riemannian connection coefficients and hence have nothing to do with torsion nor non-metricity. 

We argue that the actual purpose of teleparallel geometries is to provide a mathematical framework in which the Einstein and M\o ller Lagrangians are covariant. Therefore, without covariance, teleparallel geometries lose their main purpose. We conclude our paper by a brief discussion of modified teleparallel theories and show how they can be viewed using both approaches.	

\textit{Notation:} To distinguish between four different connections and geometric quantities related to them, we use here an extended notation of \cite{AP,Krssak:2015rqa}. The bare geometric quantities represent quantities related to a general metric-affine connection, while geometric quantities  with ``$\circ,\bullet,\raisebox{1.6pt}{\rule{4pt}{4pt}}$"  above them are related to the Riemannian, teleparallel, and symmetric teleparallel geometries, respectively. For example, $\Gamma^\rho{}_{\mu\nu},\Gammabol^\rho{}_{\mu\nu},\Gammaw^\rho{}_{\mu\nu},\Gammas^\rho{}_{\mu\nu}$ represent the linear connection of the general metric-affine, Riemannian, teleparallel, and symmetric teleparallel geometries, respectively. The Greek indices are used for the components of tensors in the coordinate basis, and the Latin indices for the components in the non-coordinate basis.

\section{Einstein's Teleparallelism and its Revival\label{secEin}}
In order to understand the origins of teleparallel gravity, we will review  Einstein's introduction of  teleparallelism\footnote{For more details and historical context of Einstein's works on teleparallelism and unified field theory, see  \cite{Sauer:2004hj,Goenner:2004se}. Another useful resource is the  English translations of the original works in German and French \cite{Delphenich}.} and the later works that lead to its revival.  In the first paper in 1928 \cite{Einstein1928a}, Einstein searched for a more general geometrical framework than the standard Riemannian geometry with the metric tensor, which could unify gravity and electromagnetism. 
He considered replacing the metric tensor by a set of four orthonormal vectors called  \textit{vierbeins} or \textit{tetrads}, representing a locally inertial reference system, which is related to the metric tensor through
\begin{equation}
	g_{\mu\nu}=\eta_{ab}h^a_{\ \mu} h^b_{\ \nu}. \label{met}
\end{equation}

The tetrad is generally not symmetric and has 16 independent components, in contrast to the 10 components of the metric tensor. It was precisely these additional 6 degrees of freedom in the tetrad that Einstein attempted to link with the 6 components of the Faraday tensor. 

To achieve this  Einstein has introduced a new geometry defined by the postulate that tetrad vectors do not rotate during the parallel transport.  This is in contrast with the Riemannian geometry where the parallel transported vectors do rotate  proportionally to the Riemannian curvature. Therefore,  Einstein has  replaced the Riemannian  covariant derivative $\nabbol_\mu$ by a new covariant derivative $\nablaw_\mu$ defined by the condition of vectors being constant during the parallel transport
\begin{equation}\label{Etelecond}
\nablaw_\mu	h_a{}^\mu=0.
\end{equation}
We can  straightforwardly solve this condition for the connection coefficients
\begin{equation}
	{\Gammawz}{}^\rho_{\ \nu\mu}=h_a^{\ \rho}\partial_\mu h^a_{\ \nu},
	\label{Weitz}
\end{equation}
where $0$ index will explained in Section~\ref{secdiff}. 
It is straightforward to  check that the curvature tensor corresponding to this connection is identically zero, but it is not symmetric, giving rise to  the torsion tensor 
\begin{equation}\label{Eintor}
\Twz{}^\rho_{\ \nu\mu}=  \Gammawz{}^\rho_{\ \nu\mu}-\Gammawz{}^\rho_{\ \mu\nu},
\end{equation}
which can be used to describe the geometry of spacetime instead of  curvature. 

In the second paper \cite{Einstein1928b},
Einstein considered an action given by the simplest scalar constructed from torsion 
\begin{equation}\label{Lag1928}
\Lw_{1928}=\frac{h}{2\kappa} \Twz^\rho{}_{\mu\nu}\Twz_\rho{}^{\mu\nu},
\end{equation} 
where $h=\det h^a_{\mu}=\sqrt{-g}$ and $\kappa=8\pi G/c^4$. Einstein then showed that perturbatively this yields the field equations of vacuum gravity and electromagnetism, where electromagnetic potential was identified with $\phi_\mu=\Twz^\nu{}_{\mu\nu}$. Einstein noted that separation of both forces is rather artificial and there is a possible non-uniqueness  in the choice of the Lagrangian \eqref{Lag1928}. This was followed by a paper by Weitzenb\"ock \cite{Weitzenbock1928}, who pointed out that the connection \eqref{Weitz} was previously considered  in \cite{Weitzenbock:1923efa}, which is why we  refer to \eqref{Weitz} as the \textit{Weitzenb\"ock connection}, and analyzed all possible invariants.  

In \cite{Einstein1929a},  Einstein considered all torsional invariants  and showed that there is an unique combination of three invariants 
\begin{equation}\label{LagTEGR}
\Lw_\text{TEGR}= \frac{h}{2 \kappa} \left(\frac{1}{4} \Twz^\rho{}_{\mu\nu}\Twz_\rho{}^{\mu\nu}+\frac{1}{2} \Twz^\rho{}_{\mu\nu} \Twz^{\nu\mu}{}_{ \rho} -\Twz^{\nu\mu}{}_{\nu} \Twz^{\nu}{}_{\mu\nu}\right),
\end{equation}
which leads to the symmetric field equations. Nowadays, we do understand that these symmetric field  are not just approximately but fully identical to the Einstein field equations, and that is why we refer to \eqref{LagTEGR} as a Lagrangian of the teleparallel equivalent of general relativity (TEGR). 

However, Einstein insisted on incorporating electromagnetism into the theory by considering a more general Lagrangian and attempted to include electromagnetism through the difference from \eqref{LagTEGR}. Einstein then again changed his approach and tried to identify electromagnetism directly with the antisymmetric part of the tetrad in the perturbative expansion  \cite{Einstein:1930xdd}. The problems started to appear, mainly due to the fact that Maxwell electromagnetism was never fully realized in these theories beyond some approximations and even Einstein kept changing his opinion how and whether it is realized in his theory. The theory was viewed critically by Eddington, Weyl and mainly Pauli, see \cite{Goenner:2004se}, and was soon abandoned.

In retrospect, it is easy to see the failure of Einstein's teleparallelism  as a consequence of a misguided motivation to identify the extra degrees of freedom of a tetrad with  electromagnetism. However, it is important to realize that Einstein's work on teleparallelism involved other important novelties, namely using tetrads instead of the metric tensor and considering other than Riemannian geometries  to describe gravity, which should not be   automatically dismissed, even if we abandon the idea of a classical unified field theory\footnote{This somewhat resembles another of Einstein's works on the cosmological constant. While the original motivation to make the Universe static was misguided, introduction of the cosmological constant has turned out to  be ultimately correct.}. 

In 1960s and 1970s it became clear that  Einstein teleparallelism can be a successful theory as long as we consider it as a theory of gravity only, i.e. without any attempt to incorporate electromagnetism. The motivation for this  came from two different directions. The first one  was M\o ller's work on the definition of the energy-momentum for gravity and second one came from the gauge approach to gravity by Hayashi, Nakano and Cho.

In order to understand both, it is useful to revisit another idea put forth by Einstein, which goes back to the very early days of general relativity. Almost immediately after introducing general relativity, Einstein became interested in a definition of the energy of gravitational field  and considered the Lagrangian  \cite{Einstein:1916cd}
\begin{equation}\label{LagEin}
	\Lag_\text{E}= \frac{1}{2\kappa}\sqrt{-g}g^{\mu\nu}(
	\Gammabol^\rho{}_{\sigma\mu}\Gammabol^\sigma{}_{\rho\nu}-
	\Gammabol^\rho{}_{\mu\nu}\Gammabol^\sigma{}_{\rho\sigma}
	),
\end{equation}
from which  the gravitational energy-momentum was defined as
\begin{equation}\label{Einsteinpseudo}
t^\mu{}_\nu=\Lag_\text{E} \, \delta^\mu{}_\nu- \partial_\nu g^{\rho\sigma}\frac{\partial \Lag_\text{E}}{\partial_\mu g^{\rho\sigma}},
\end{equation}
which became known as the Einstein gravitational energy-momentum pseudotensor, and  was followed by introduction other similar pseudotensors, e.g. the symmetric one by Landau and Lifshitz \cite{Landau:1982dva}. All these pseudotensors have a rather  undesirable property that they need to be  evaluated in  well-behaved coordinate systems, which asymptotically approach the inertial coordinate system \cite{Einstein:1918,Landau:1982dva}, but they do give an acceptable definition of the gravitational energy-momentum \cite{Chang:1998wj,Chen:2018geu}. 

The work of M\o ller \cite{Moller1961} was motivated by finding a well-behaved tensorial object to describe the gravitational energy-momentum. His solution was to use tetrads instead of the metric to describe gravity, but tetrads were considered still within the Riemannian geometry.  The  M\o ller Lagrangian is expressed in terms of the Ricci coefficients of rotation, see Section~\ref{secgeom}, as \cite{Moller1961}\footnote{M\o ller originally wrote this using the coordinate indexed connection $\ombol^\rho{}_{\mu\nu}=h_a{}^\rho h^b{}_\mu h^c{}_\nu\ombol^a{}_{bc}$. We will return to this in Section~\ref{secdiff}. }
\begin{equation}\label{LagMoll}
	\Lag_\text{M}=	 \frac{1}{2\kappa}h 
	\left(\ombol^{a}_{\,\,\,ca}\ombol^{bc}{}_{b}
	-\ombol^{a}{}_{cb}
	\ombol^{bc}{}_{a} 
	\right),
\end{equation}
and we can define the M\o ller energy-momentum complex as
\begin{equation}\label{Mollercomplex}
	\tilde{t}^\mu{}_\nu=\Lag_\text{M} \, \delta^\mu{}_\nu- \partial_\nu h_a{}^\rho\frac{\partial \Lag_\text{M}}{\partial_\mu h_a{}^\rho},
\end{equation}
which was argued to be a tensor with respect to the coordinate transformations and hence was claimed to solve the main problem of Einstein's pseudotensor \eqref{Einsteinpseudo}.

However, in order to provide a meaningful result, it had to to be supplemented by six conditions on the tetrad 
\begin{equation}\label{Mollercond}
\phi_{ab}=0,
\end{equation}
where $\phi_{ab}$ is antisymmetric object constructed from the tetrad and its derivatives. These were originally determined in the weak field limit, and in our notation take the form as $\phi_{ab}=\ombol_{acb}\,\ombol^{dc}{}_{d}$. The problem of finding the well-behaved coordinate system for \eqref{Einsteinpseudo} was then turned into a problem of finding six conditions on the tetrad, which was further discussed in \cite{Moller1966,Pellegrini1963,Moller1978}.

While the starting point  of  M\o ller was the Riemannian geometry, and both \eqref{Mollercomplex} and \eqref{Mollercond} were derived within the Riemannian geometry, at the end of the paper M\o ller noticed that the condition \eqref{Mollercond} determines the orientation of the tetrads that are fixed throughout the whole spacetime. He argued that this is  the case of the Weitzenb\"ock geometry \eqref{Weitz} and  proceeds to discuss Einstein's teleparallelism. Therefore,  the M\o ller's work \cite{Moller1961} is generally viewed as the first paper  since 1930s that acknowledges teleparallelism as a viable theory.

The second motivation to revisit Einstein's teleparallelism came from works on the gauge aspects of gravity. The first paper to discuss gravity as a gauge theory of translations only was by Hayashi and Nakano 
\cite{Hayashi:1967se}, who argued in its favor over gauging the Lorentz or Poincare groups \cite{Blagojevic:2012bc}. They identified the torsion tensor \eqref{Eintor} as the fields strength of the translational group, but considered a generalized Lagrangian that later became known as \textit{new general relativity} \cite{Hayashi:1979qx}. It was Cho \cite{Cho:1975dh} who first argued that it is possible to view general relativity as a gauge theory of translations, if we require the covariance of the field equations, which restricts the form  of the translational Lagrangian to the Einstein's  teleparallel Lagrangian \eqref{LagTEGR}. 

It is interesting to note that Cho \cite{Cho:1975dh} never explicitly mentions Einstein's teleparallelism. He stated that his translational Lagrangian is up to a divergence equivalent to the Einstein Lagrangian, from where it seems that he meant rather an equivalence with the  Einstein-Hilbert Lagrangian.
The fact that the underlying geometry is the Weitzenb\"ock geometry seems to be first realized by Hayashi in \cite{Hayashi:1977jd}, where Hayashi  argued against Cho's requirement for the covariant field equations in favor of the generalized Lagrangian \cite{Hayashi:1979qx}.

\section{Metric-Affine Approach to Teleparallel Geometries\label{secgeom}}
In the original Einstein's teleparallelism described in the previous section, the distant parallelism condition  \eqref{Etelecond} was introduced as a postulate of the new geometry. In this section, we will present a more general viewpoint where teleparallel geometries are viewed as  special cases of   general metric-affine geometries \cite{Obukhov:2002tm}. Along the way, we will briefly review  some elements of differential geometry essential for our discussion. For more details, see \cite{Nakahara:2003nw,Fecko:2006zy,Hehl:1994ue}. 

We are interested in studying the geometry of a differentiable manifold $\Mag$, which is a  topological space on which we can define local coordinates. At each point $p$, we can then define the tangent space 
and its dual cotangent space, elements of which are vectors and covectors, from  where we define then the vector and covector fields.
The coordinates define a coordinate basis $\{\partial_\mu\}$ for the vector fields and $\{dx^\mu\}$ for 1-forms or covector fields, which can be written as 
\begin{equation}\label{vectors}
V=V^\mu\partial_\mu, \qquad \omega=\omega_\mu dx^\mu,	
\end{equation}
where $V^\mu$ and $\omega_\mu $ are components of the vector $V$ and covector $\omega$ in the coordinate basis. A general tensor field  can be then written in the coordinate basis as
\begin{equation}\label{ten}
X=X^{\mu_1\dots\mu_p}{}_{\nu_1\dots\nu_g}
\partial_{\mu_1}\otimes\dots\otimes\partial_{\mu_p}\otimes dx^{\nu_1}\otimes\dots \otimes dx^{\nu_q}.	
\end{equation} 
The tensor fields are invariant under a change of coordinates, and  we can determine the  transformation properties of tensor components from the knowledge of  coordinate basis transformations. Under the coordinate change, $x^\mu\rightarrow x^{\mu'}$, the tensor components transform as
\begin{equation}
	X^{\mu'\dots}{}_{\nu'\dots}=X^{\mu\dots}{}_{\nu\dots}\, \frac{\partial x^{\mu'}}{\partial x^\mu}\dots \frac{\partial x^\nu}{\partial x^{\nu'}}\dots, \label{tencoordtransf}
\end{equation}

We can consider a more general class of bases by taking $4$ independent vectors that we  denote $h_a$, where $a=1\dots 4$, and use them as a basis for the vector fields in the tangent space. Equally, we can consider  dual covectors  $h^a$ as a basis in the cotangent space,  $h^a h_b=\delta^a_b$.
These basis  vectors and covectors can be  expressed in the coordinate basis as
\begin{equation}\label{key}
h^a=h^a_{\ \mu}dx^\mu,\qquad h_a=h_a{}^\mu \partial_\mu.	
\end{equation}

If these general basis vectors do not commute, i.e. $[h_a,h_b]\neq 0$, it is not possible to express them in terms of some new coordinates, and  we say that $h_a$ is  a \textit{non-coordinate basis}.
We can then
write an arbitrary tensor field in the non-coordinate basis, and obtain a relation between components of a tensor in the coordinate and  non-coordinate basis as
\begin{equation}
	X^{a\dots}{}_{b\dots}=X^{\mu\dots}{}_{\nu\dots}\, h^a_{\ \mu}\dots h_b^{\ \nu}\dots  \label{tetproj}
\end{equation}

In the non-coordinate basis,  the components of a tensor are invariant under a coordinate change since  $h^a_{\ \mu}$ transforms as a coordinate covector in the last index. We can instead consider an arbitrary change of basis as
\begin{equation}\label{basistransf}
h^{a'}=\Lambda^{a'}{}_b h^b,
\end{equation}
where $\Lambda^{a'}{}_b$ is a general non-singular matrix, i.e. $\Lambda^{a'}{}_b \in GL(4)$. We often write this as $\Lambda^{a}{}_b$, i.e. $h^{a'}=\Lambda^{a}{}_b h^b$. The  tensor components in the non-coordinate basis do change under \eqref{basistransf} as
\begin{equation}
	X^{a'\dots}{}_{b'\dots}=X^{a\dots}{}_{b\dots}\, \Lambda^{a'}{}_a\dots (\Lambda^{-1})^b{}_{b'}\dots. \label{tenanholotrans}
\end{equation}

As the next step, we introduce the metric tensor on a manifold that allows us to measure distances and angles, leading us to the \textit{Riemmanian manifold} $(\Mag,g)$\footnote{To be more precise, we are interested in the  pseudo-Riemannian or Lorentzian manifolds.
}. The components of the metric tensor in the non-coordinate basis are then 
\begin{equation}
g_{ab}=g_{\mu\nu}h_a^{\ \mu} h_b^{\ \nu}. \label{met1}	
\end{equation}

The interesting case for us, is the class of orthonormal non-coordinate bases, i.e. the case where we fix the metric in the tangent space to be the Minkowski metric, $g_{ab}=\eta_{ab}=\text{diag}(-1,1,1,1)$, which  we call the \textit{tetrads}\footnote{As is common in the literature, the term tetrad we use for both  $h^a$ and $h_a$, as well as their components. 
}. Then the components of the metric tensor $g_{\mu\nu}$, can be expressed from \eqref{met1} as
\begin{equation}
	g_{\mu\nu}=\eta_{ab}h^a_{\ \mu} h^b_{\ \nu}, \label{met2}
\end{equation}
which is precisely the relation considered by Einstein \eqref{met}.

We would like to highlight here that it is this restriction to the class of orthonormal non-coordinate bases that  allows us to use the tetrad instead of the metric as a fundamental variable\footnote{Note that sometimes this is presented as ``differential geometry without a metric" \cite{Stephani:2003tm}, but we should keep in mind that we have the metric $\eta_{ab}$ in the tangent space, but we keep it fixed and hence  non-dynamical.}. %
Moreover, this narrows down the transformations of the basis \eqref{basistransf} to those that preserve the condition of orthonormality, i.e. such transformations that leave the tangent space metric $\eta_{ab}$ invariant, from where immediately follows  that  $\Lambda^{a}{}_b$ must be a local Lorentz transformation, i.e.  $\Lambda^{a}{}_b \in SO(1,3)$.

The last step involves introducing the \textit{covariant derivative} $\nabla$ on the Riemannian  manifold  $(\Mag,g)$. This is motivated by the fact that  the ordinary partial derivative of a tensor field does not yield a tensor field. To address this, we essentially define the covariant derivative in a way that ensures the resulting object is  a tensor field and acts as a derivative.  We can  characterize the covariant derivative by the connection coefficients, and since we consider both coordinate and non-coordinate bases, it is useful to distinguish between the covariant derivative of components of tensors in different  bases, and introduce two  kinds of connection coefficients
\begin{equation}
	{\nabla}_\nu X^\rho=\partial_\nu X^\rho + {\Gamma}^\rho{}_{\mu\nu}X^\nu, \qquad
	{\nabla}_\nu X^a=\partial_\nu X^a + {\omega}^a{}_{b\nu}X^b,\label{covdef}
\end{equation} 
where we call  ${\Gamma}^\rho{}_{\mu\nu}$ the  \textit{linear connection coefficients}, and ${\omega}^a{}_{b\nu}$  the \textit{spin connection coefficients} \cite{AP}. 

However, it is important to understand that these are merely two different connection coefficients for  the same covariant derivative in two different bases. For arbitrary vector fields  $X=X^\mu \partial_\mu= X^a h_a$ and $Y=Y^\mu\partial_\mu$, the covariant derivative of a vector in the coordinate basis $\nabla_Y X=(Y^\nu \nabla_\nu X^\mu) \partial_\mu $ must be then same as the covariant derivative of the same vector in non-coordinate basis $\nabla_Y X= (Y^\nu \nabla_\nu X^a) h_a$, which immediately gives us a relation between the linear and spin connection coefficients
\begin{equation}
	\Gamma^\rho_{\ \nu\mu}=h_a^{\ \rho}\partial_\mu h^a_{\ \nu}+
	h_a^{\ \rho} \omega^a_{\ b\mu}h^b_{\ \nu}.
	\label{Chris}
\end{equation}

The reason why it is useful to keep this distinction between the linear and spin connection coefficients is due to different transformation properties of tensor components in the coordinate and non-coordinate bases, from where the connection coefficients inherit their transformation properties. 
Under the coordinate change $x^\mu\rightarrow x^{\mu'}$, the tensor components transform as \eqref{tencoordtransf}, and hence the linear connection transforms as
		\begin{equation}\label{gammatransf}
	\Gamma^{\rho'}{}_{\mu'\nu'}=
	\frac{\partial x^\mu}{\partial x^{\mu'}}
	\frac{\partial x^\nu}{\partial x^{\nu'}}
	\frac{\partial x^{\rho'}}{\partial x^{\rho}}
	\Gamma^{\rho}{}_{\mu\nu}
	-\frac{\partial x^\mu}{\partial x^{\mu'}}
	\frac{\partial x^\nu}{\partial x^{\nu'}}
	\frac{\partial^2 x^{\rho'}}{\partial x^\mu \partial x^\nu}
\end{equation}
while the spin connection does transform as a tensor in the last index.

Under the change of non-coordinate basis \eqref{basistransf}, the components of tensors in a  non-coordinate basis transform as \eqref{tenanholotrans}, and hence the spin connection transforms as
\begin{equation}\label{omegatransf}
	\omega^{a'}_{\ b'\mu}=\Lambda^{a}_{\ c}{\omega}^c_{\ d\mu}\Lambda_b^{\ d}+\Lambda^a_{\ c} 
	\partial_\mu (\Lambda^{-1})^c{}_b,
\end{equation}
while the linear connection is invariant under such a change of basis.

In the most general case,  both connection coefficients  have $4^3=64$ independent components. We can characterize the connection in terms of three tensors known as: curvature, torsion and non-metricity. It is again useful to distinguish between working in the coordinate  and non-coordinate bases. In the coordinate basis, it is natural to work in terms of the pair $\{g_{\mu\nu},\Gamma^\rho{}_{\mu\nu}\}$ and we define these three tensors as 
\begin{eqnarray}\label{curvhol}
 		R^\mu{}_{\nu\rho\sigma} &\equiv&\partial_\rho
 		\Gamma^\mu{}_{\nu\sigma}-
	\partial_\sigma \Gamma^\mu{}_{\nu\rho}
	+\Gamma^\tau{}_{\nu\sigma}\Gamma^\mu{}_{\tau\rho}-
	\Gamma^\tau{}_{\nu\rho}\Gamma^\mu{}_{\tau\sigma},\\
	T^\rho{}_{\mu\nu}&\equiv& \Gamma^\rho{}_{\mu\nu}-\Gamma^\rho{}_{\nu\mu},\\
	Q_{\rho\mu\nu}&\equiv&\nabla_\rho g_{\mu\nu}.
\end{eqnarray}

In the non-coordinate basis, it is natural  to use the tetrad and the spin connection, i.e. work with the pair $\{h^a{}_{\mu},\omega^a{}_{b\mu}\}$, and define  these  tensors as\footnote{We follow here the formalism used in \cite{AP}, which differentiates between the two groups of indices explicitly to stress that these tensors are  components of the Lie algebra-valued differential forms. }
\begin{eqnarray}
		R^{a}_{\,\,\, b\mu\nu} &\equiv& \partial_\mu\omega^{a}_{\,\,\,b\nu}-
	\partial_\nu\omega^{a}_{\,\,\,b\mu}
	+\omega^{a}_{\,\,\,c\mu}\omega^{c}_{\,\,\,b\nu}-\omega^{a}_{\,\,\,c\nu}
	\omega^{c}_{\,\,\,b\mu},
	\label{curv}\\
	T^a_{\ \mu\nu}&\equiv&\partial_\mu h^a_{\ \nu} -\partial_\nu h^a_{\ \mu}+\ombol^a_{\ b\mu}h^b_{\ \nu}
-\ombol^a_{\ b\nu}h^b_{\ \mu}, \label{torsion} \\
Q_{\mu a b}&\equiv& \nabla_\mu \eta_{ab}=\omega_{ab\mu}+\omega_{ba\mu}.\label{nonmet}
\end{eqnarray}

The most general geometries with 64 independent components of the connections are known as the \textit{metric-affine geometries} \cite{Hehl:1994ue}.
For the reasons that will be explained in detail in Section~\ref{secmeaning}, we primarily focus on three  limiting cases defined by taking two of these three tensors to zero.

\subsubsection*{Riemannian connection}
The most well-known geometry is the Riemannian or Levi-Civita geometry defined by  conditions of vanishing torsion and non-metricity. In the coordinate basis, we solve these conditions to find that the Riemannian linear connection is uniquely determined in terms of the metric tensor as
\begin{equation}\label{Christoffel}
	\Gammabol^\rho{}_{\nu\mu}=\frac{1}{2} g^{\rho\sigma}\left(
	g_{\nu\sigma,\mu} + 
	g_{\mu\sigma,\nu}-
	g_{\nu\mu,\sigma}
	\right).
\end{equation}
In the non-coordinate basis, we find an analogous result that the Riemannian spin connection is determined fully in terms of the tetrad as
\begin{equation}
	\ombol^a{}_{b\mu}=\frac{1}{2} h^{\;c}{}_{\mu} \Big[f_b{}^a{}_c + f_c{}^a{}_b - f^a{}_{bc}\Big],\label{lccon}
\end{equation}
where 
\begin{equation}\label{anholcoef}
		f^c{}_{a b} = h_a{}^{\mu} h_b{}^{\nu} (\partial_\nu
	h^c{}_{\mu} - \partial_\mu h^c{}_{\nu} ),
\end{equation}
are the coefficients of anholonomy. We can then define also $\ombol^a{}_{bc}=\ombol^a{}_{b\mu} h_a{}^\mu$ or $\ombol^\rho{}_{\nu\mu}=h_a{}^\rho h^b{}_\nu \ombol^a{}_{b\mu}$ that are often called the \textit{Ricci coefficients of rotation}.

\subsubsection*{Teleparallel connection}
The teleparallel connection, occasionally referred as the metric teleparallel connection, is defined by the so-called teleparallel condition, i.e. vanishing curvature, and metric-compatibility. The zero-curvature condition is solved by a pure gauge connection
\begin{equation}
	\omw^a_{\ b\mu}=\Lambda^a_{\ c} \partial_\mu (\Lambda^{-1})^c{}_b,
	\label{telcon}
\end{equation}
where $\Lambda^a_{\ b}$ is a non-singular matrix. The metric compatibility connection then restricts $\Lambda^a_{\ b}$ to be a local Lorentz transformation, i.e. $\Lambda^a_{\ b}\in SO(1,3)$. 
The corresponding linear teleparallel connection is then
\begin{equation}
	\Gammaw^\rho_{\ \nu\mu}=h_a^{\ \rho}\partial_\mu h^a_{\ \nu}+
	h_a^{\ \rho} \omw^a_{\ b\mu}h^b_{\ \nu}.
	\label{Christele}
\end{equation}

The crucial relation is the Ricci theorem that allows us relate the teleparallel spin connection \eqref{telcon} and the Riemannian spin connection \eqref{lccon}  as 
\begin{equation}\label{Ricci}
	\omw^a_{\ b\mu}=\ombol^a_{\ b\mu} + \Kw^a_{\ b\mu},
\end{equation}
where 
\begin{equation}
	\Kw^{a}_{\  b\mu}=\frac{1}{2}
	\left(
	\Tw^{\ a}_{\mu \ b}
	+\Tw^{\ a}_{b \ \mu}
	-\Tw^{a}_{\  b\mu}
	\right),
	\label{contortion}
\end{equation} 
is the contortion tensor, and an analogous relation holds for \eqref{Christele} and \eqref{Christoffel}.

\subsubsection*{Symmetric teleparallel connection}
Analogously to the previous case, we can consider the symmetric teleparallel geometry, where the connection is defined by the teleparallel condition and the condition of being symmetric. This geometry was first considered by Nester and Yo \cite{Nester:1998mp} and later explored in  \cite{BeltranJimenez:2017tkd,BeltranJimenez:2018vdo}.

The teleparallel condition has a similar pure gauge-type solution as \eqref{telcon} and together with the symmetricity condition determines the symmetric teleparallel linear connection to be given by \cite{BeltranJimenez:2018vdo}
\begin{equation}\label{symtelcon}
\Gammas^\rho{}_{\mu\nu}=
\frac{\partial x^\rho}{\partial \xi^\alpha}\frac{\partial}{\partial_\mu} \frac{\partial \xi^\alpha}{\partial x^\nu},	
\end{equation}
where $\xi^\alpha$ are arbitrary four functions.  We can then find the analogue of the Ricci theorem \eqref{Ricci} as
\begin{equation}\label{key}
\Gammas^\rho{}_{\mu\nu}=
\Gammabol^\rho{}_{\mu\nu}+
\Ls^\rho{}_{\mu\nu}	
\end{equation}
where $\Ls^\rho{}_{\mu\nu}=\frac{1}{2}\left(
\Qs^\rho{}_{\mu\nu}-\Qs_\mu{}^\rho{}_{\nu}
-\Qs_\nu{}^\rho{}_{\mu}
\right)$ is the disformation tensor.
\section{Einstein's Teleparallelism and Teleparallel Gravity\label{secdiff}}
We can observe that we have in fact two distinct definitions of teleparallel geometries. In the original Einstein's teleparallelism, the geometry is defined by the condition  
\eqref{Etelecond} that leads to  zero curvature. In the covariant or metric-affine approach, we  define  teleparallel geometry  by the condition of zero curvature and obtain a more general teleparallel spin connection \eqref{telcon}.  

It is straightforward to relate both approaches following the fact that the teleparallel spin connection \eqref{telcon} is of a pure gauge form, which means that using \ref{omegatransf} we can always find a local Lorentz transformation that transforms this connection to zero
\begin{equation}\label{gaugeWeitz}
	\omwz^a{}_{b\mu}=0,
\end{equation}
which is referred as the Weitzenb\"ock gauge \cite{Obukhov:2002tm}. 
We can consider then  the Weitzenb\"ock connection \eqref{Weitz} as just the universal term in \eqref{Chris} responsible for the relation between coordinate and non-coordinate basis, and hence \eqref{Weitz}  is just the linear connection corresponding to the zero spin connection \cite{deAndrade:2001vx}, from where should be clear why we added the index 0 in Section~\ref{secEin}.

This is indeed a subtle difference, but there are  serious consequences as far as the mathematical consistency of teleparallel theories is concerned. We can  see that the ``torsion" in the original Einstein's teleparallelism \eqref{Eintor} is not actually a tensor, but just the coefficient of anholonomy
\begin{equation}\label{faketor}
	\Twz{}^\rho_{\ \nu\mu}=f^\rho_{\ \nu\mu}.
\end{equation}

Our argument is that this represents a serious mathematical consistency problem since non-tensorial quantities are being erroneously called tensors. Nevertheless,  despite this problem, \eqref{faketor} is sufficient to establish the equivalence between \eqref{LagMoll} and \eqref{LagTEGR}, by straightforwardly substituting \eqref{faketor} into \eqref{Ricci}. We get
\begin{equation}\label{Ricciwrong}
	\ombol^\rho{}_{\mu\nu}= - \Kwz^\rho{}_{\mu\nu},
\end{equation}
where $\Kwz^\rho{}_{\mu\nu}$ is  the contortion \eqref{contortion} obtained using \eqref{faketor}.

However, unlike the actual Ricci theorem \eqref{Ricci}, which makes a sound mathematical statement  that a difference between two connections is a tensor, now the Riemannian connection is misidentified with the contortion tensor \eqref{Ricciwrong}.  Our main argument in this paper is that  we should either use the Riemannian connection or work with the actual teleparallel tensors, but we should never make identifications of the kind \eqref{Ricciwrong} or \eqref{faketor}. We do believe that this is actually the source of the confusion about the foundational issues of teleparallel theories. 

We note here that in the discussion about whether some object is a tensor or not, the crucial factor is not whether it has  Greek or Latin indices, but  how it depends on the choice of a basis.
For example, objects like  $f^\rho_{\ \nu\mu}$ or $\ombol^\rho{}_{\mu\nu}$ are  formally  ``coordinate-tensors", but this is actually irrelevant. These objects are not true tensors since they depend on the choice of a non-coordinate basis in a non-tensorial way. Indeed, these are  just the coefficients of anholonomy and Ricci rotation coefficients, and there is no benefit calling them torsion and  contortion tensors.

The same applies to the  M\o ller complex \eqref{Mollercomplex}, which is formally a ``coordinate-tensor" but depends on the  non-coordinate basis in a non-tensorial way. This is why the choice of the tetrad must be fixed using the additional conditions \eqref{Mollercond}.  
We can  observe that the M\o ller Lagrangian \eqref{LagMoll} and complex \eqref{Mollercomplex} are just straightforward tetrad analogues of Einstein Lagrangian \eqref{LagEin} and pseudotensor \eqref{Einsteinpseudo}. The difficulty of finding a well-behaved coordinate system for the Einstein pseudotensor is  replaced by finding a well-behaved tetrad for the M\o ller complex.


On the other hand, the actual teleparallel torsion tensor is
\begin{equation}\label{teletor}
	\Tw{}^\rho{}_{\nu\mu}=f^\rho{}_{\nu\mu}+ \omw^\rho{}_{\nu\mu}-\omw^\rho{}_{\mu\nu},	
\end{equation}
where the presence of the teleparallel spin connection is crucial to make it a proper tensor and hence to be called the torsion tensor. The Lagrangian of teleparallel equivalent of general relativity is then
\begin{equation}\label{LagTEGRcov}
	\Lw_\text{TEGR}= \frac{h}{2 \kappa} \left(\frac{1}{4} \Tw^\rho{}_{\mu\nu}\Tw_\rho{}^{\mu\nu}+\frac{1}{2} \Tw^\rho{}_{\mu\nu} \Tw^{\nu\mu}{}_{ \rho} -\Tw^{\nu\mu}{}_{\nu} \Tw^{\nu}{}_{\mu\nu}\right),
\end{equation}
which is  a proper scalar (density) under a change of  both coordinate and non-coordinate bases.

This brings us to the main question what problems the covariant approach actually solves. The mathematical consistency issues of these theories is indeed solved, i.e. it makes all the objects we refer to as tensors actual tensors.  Moreover, it allows us to use arbitrary tetrads and hence avoids the issue of working in the class of preferred tetrads. As we will argue in Section~\ref{secmeaning}, both these aspects are realized naturally  using only the existing structure on the manifold.

However, it is important to realize that all these features are achieved by determining the spin connection and matching it with the tetrad, what makes the teleparallel gravity computationally equivalent to the M\o ller approach.  While in the the M\o ller gravity we have to determine a preferred class of tetrads by finding $\Lambda^a{}_b$ that transform to the preferred class from a generic tetrad, in teleparallel gravity we use the same $\Lambda^a{}_b$ to calculate the spin connection \eqref{telcon}, which  we  then associate with the tetrad.  
 
The procedure to determine the spin connection is still not fully understood and some unresolved problems remain. The first time the non-trivial spin connection was used in works of Obukhov and Pereira \cite{Obukhov:2002tm,Lucas:2009nq}, who found it as the asymptotic limit of the Riemannian connection in the Schwarzschild case and demonstrated the regularizing properties on conserved charges. In \cite{Krssak:2015rqa, Krssak:2015lba}, it was argued that the underlying principle to determine the spin connection should be the finiteness of the action, and that the teleparallel spin connection can be always calculated as
\begin{equation}
	\ombol^a{}_{b\mu} (h_{(\rm r)}^{\;a}{}_{\mu}), 
	\label{regconsolution2}
\end{equation}
where $h_{(\rm r)}^{\;a}{}_{\mu}$  is the reference tetrad, i.e. some tetrad in the Minkowski spacetime that we associate with the ``full" tetrad. This follows from a simple observation that in the Minkowski spacetime the most general tetrad is given by a local Lorentz transformation of an inertial tetrad, i.e. the tetrad for which the Riemannian connection \eqref{lccon} vanishes, and hence the Riemannian connection of the general Minkowski tetrad is guaranteed to have the pure gauge form \eqref{telcon}. 

It was argued that  this method can be viewed as an analogue of the ``background subtraction" method of Gibbons and Hawking \cite{Gibbons:1976ue}, where we do obtain a finite action using a reference metric as well, i.e. that the teleparallel action is equivalent to the full action including the Gibbons-Hawking boundary term \cite{Krssak:2015rqa}.  Recently \cite{Krssak:2023nrw}, we have argued that the same result can be achieved in the interior region of a black hole, where we obtain a perfect agreement with the Gibbons-Hawking boundary term even for the asymptotically AdS black holes.  

Nevertheless, we still do lack an universal general method of determining spin connection. As it was pointed out in \cite{Maluf:2018coz}, there seems to be some non-uniqueness since multiple tetrads can be associated with the same spin connection, which shows that we lack a completely general principle how to match the reference tetrad with the full tetrad. In particular, this means that the so-called ``switching-off gravity" method to determine the reference tetrad in \cite{Krssak:2015rqa,Krssak:2015oua,Krssak:2018ywd} should not be viewed as a general method but rather a tool with a limited use in some special cases \cite{Emtsova:2021ehh}\footnote{Another method of determining the spin connection is based on the symmetry arguments, where we require the connection to have the same symmetries as the tetrad \cite{Hohmann:2019nat}.}. However, we should remember that there is an analogous issue in the background subtraction method, which is known to not work in a completely general case, so perhaps one should not expect a too simple solution to a very complex problem of regularization of the gravitational action.

However, we would like to stress here that even  if there is some ambiguity in determining the teleparallel spin connection, this is an ambiguity in determining $\Lambda^a{}_b$. Therefore, we have a completely analogous ambiguity in the M\o ller gravity or the non-covariant teleparallel gravity, where this ambiguity  will demonstrate itself through existence of multiple non-equivalent  proper tetrads for the same spacetime \cite{Emtsova:2021ehh}.

We can observe that  our discussion essentially  goes back to the Einstein  Lagrangian \eqref{LagEin} and represents various reformulations and improvements in dealing  with it. The problem of finding the preferred coordinate basis for evaluating the  Einstein action \eqref{LagEin} and pseudotensor \eqref{Einsteinpseudo}, was recast by M\o ller into finding a preferred non-coordinate basis for evaluating \eqref{LagMoll} and \eqref{Mollercomplex}. The covariant teleparallel gravity then translates this issue into a question of how to associate the teleparallel spin connection (or a reference tetrad) with the tetrad. We can see the fundamental ``law of conservation of difficulty" is indeed at play, and the steps towards improvement are rather incremental. Nevertheless,  from this perspective, M\o ller's work improved the situation by allowing us to use arbitrary coordinates, and the covariant teleparallel gravity is a further improvement that  enabled us to use arbitrary tetrads.

\section{Geometrical Significance of Teleparallelism\label{secmeaning}}
The Riemannian connection is usually seen as the preferred connection on the Riemannian manifold.  This follows directly from the fact that it is induced by the metric and hence the Riemannian manifold $(\Mag,g)$ comes naturally with the Riemannian connection. This property allows us to take Riemannian covariant derivatives without any additional structure and talk about the curvature of the Riemannian manifold or a metric on its own. We would like to show here that teleparallel geometries are equally privileged in the sense that they are  defined without any additional structure, and that they are fundamentally different from the general metric-affine geometries. 

In  metric-affine geometries, we  cannot meaningfully talk about the curvature of a metric, since curvature is not a property of the metric itself, but rather of the connection \eqref{curvhol}. This means that  assigning curvature to any metric is not possible without the connection that is \textit{a priori} independent from the metric.  We then have to either work with the  pair of variables $\{h^a{}_{\mu},\omega^a{}_{b\mu}\}$\footnote{Or equivalently, with the pair $\{g_{\mu\nu},\Gamma^\rho{}_{\mu\nu}\}$, depending on whether we use the coordinate or non-coordinate basis. From now on, for simplicity, we focus on the Einstein-Cartan geometry as a representative of metric-affine geometries,  and comment only if there is some difference in the case of non-metricity.},  or use the Ricci theorem \eqref{Ricci} to decompose them into the Riemannian connection and the  contortion tensor.

There are then two very different school of thoughts how to proceed, as was  illustrated in the short but interesting discussion  between  Weinberg and Hehl
\cite{Hehl}. While Weinberg  argued that geometry is given by the Riemannian connection and torsion  is ``just a tensor" that can be treated as an additional (matter) field in the context of general relativity,  Hehl contended that torsion is a  very specific tensor tied to the very geometric structure of the manifold and translation group. No matter which viewpoint we choose to follow,  in generalized geometries like the  Einstein-Cartan one, the independent  connection does  add extra independent degrees of freedom, which represent deviation from the Riemannian geometry and this deviation is expected to be measurable \cite{Mao:2006bb}.

The situation is  different in the case of teleparallel geometries, since the teleparallel torsion is not considered as some additional field to the Riemannian connection or its curvature, but it rather ``takes the role of curvature", in the sense that  we use it to describe the non-trivial structure of spacetime instead of curvature. This means that unlike the Einstein-Cartan torsion, the teleparallel  torsion is measured whenever we observe some effect of curvature in general relativity. The same argument can be made about non-metricity in the symmetric teleparallel case.

Moreover, both teleparallel geometries  can be defined on the Riemannian manifold without any additional structure. This naively seems to go against the traditional viewpoint that only the Riemannian connection is defined on the Riemannian manifold, and everything else must be added as an additional structure, but we should keep in mind that the manifold by definition comes with the existence of a basis. Teleparallel geometries can be then considered to be unique geometries where the connection carries information about the basis only.  In the case of torsional teleparallel geometry, the connection \eqref{telcon} is fully given by local Lorentz transformations $\Lambda^a{}_b$, and in the symmetric teleparallel case, the connection \eqref{symtelcon} is fully given by the choice of coordinates. 

Now comes the main difference with the Riemannian case: since the basis is not unique, each choice of a basis yields a different connection. As the torsion \eqref{torsion} depends on both the tetrad and connection, each choice of a basis will not only produce a different connection but also a different torsion tensor. Therefore, both the teleparallel connection and torsion are not uniquely determined. This contrasts with the Riemannian case, where all the information about the geometry of the manifold, including the properties of the basis, is encoded in a single quantity—the Riemannian connection.

In teleparallel geometry, we have naturally two quantities and characterize the geometry in terms of both the teleparallel connection and torsion\footnote{We can choose to use any of the following equivalent pairs of the variables: torsion and teleparallel connection, tetrad and teleparallel connection, or Riemannian connection and contortion.}. It is important to realize that together they carry the same information as the Riemannian connection, simply by the virtue of Ricci theorem \eqref{Ricci}, which states that together they give the Riemannian spin connection 
\begin{equation}\label{Riccisplit}
	\ombol^a_{\ b\mu} = \omw^a_{\ b\mu}-  \Kw^a_{\ b\mu}.
\end{equation}

Therefore,  we are guaranteed to not loose any information about the manifold. 
We can view \eqref{Riccisplit} as that each choice of a basis, defines a split  of the Riemannian connection into a  connection  piece that carries information about the basis only and the contortion tensor. As a consequence, we naturally obtain the contortion as a tensorial field strength of gravity\footnote{Note that in the literature, including \cite{AP,Krssak:2018ywd}, it is possible to find a claim that contortion is a purely gravitational strength, i.e. absent of any inertial effects. More careful analysis in our upcoming paper \cite{Krssak1}, demonstrates that this claim is not entirely correct. However, for our discussion here,  it is relevant that contortion is a true tensor.}, and hence the Lagrangian \eqref{LagTEGRcov} is a proper scalar density. 

The most important consequence from this viewpoint is that it establishes a relation between the tetrad and teleparallel connection. This is a crucial distinction from metric-affine geometries, where the tetrad and spin connection are fully independent variables. We can now interpret teleparallel geometry as that we first calculate the Riemannian spin connection from the tetrad alone, and then the choice of $\Lambda^a{}_b$ defines its split into the teleparallel connection and contortion \eqref{Riccisplit}. Consequently, we find that $\Lambda^a{}_b$ in the connection \eqref{telcon} and $\Lambda^a{}_b$ used in the torsion \eqref{teletor} are the same $\Lambda^a{}_b$, which is equivalent to the statement that the tetrad and teleparallel spin connection are not independent variables. 

A completely analogous situations is in the symmetric teleparallel case, where the choice of a basis is related to the choice of the coordinates, and non-metricity plays the role of torsion. Then each choice of the coordinates defines a split of Riemannian linear connection into a symmetric teleparallel linear connection and disformation tensor proportional to the non-metricity. We analogously do have the non-metricity tensor as a true tensorial object that plays the role of a gravitational field strength.

\section{Conclusions\label{secconc}}
In recent years, the topic of covariance of teleparallel theories was a subject of various  discussions that essentially concern the treatment of the teleparallel  spin connection, and whether the so-called Wietzenb\"ock gauge \eqref{gaugeWeitz} is the fundamental definition of teleparallel theories or one should use the non-trivial spin connection. The same kind of discussion can be made in the case of symmetric teleparallel theories, where the question is about the coincident gauge.

We have argued here in favor of the so-called  covariant  approach to teleparallel gravity theories  based on the metric-affine construction presented in section~\ref{secgeom}, where the condition of vanishing curvature results in the pure gauge connection \eqref{telcon}. The  Lagrangians of teleparallel models constructed in this way are invariant under simultaneous local Lorentz transformations of the tetrad and teleparallel spin connection.
The main advantage 
is from the viewpoint of mathematical consistency, as it guarantees that all quantities used in teleparallel theories are actual tensors, and we can perform calculations using an arbitrary basis and avoid working in a preferred class of tetrads.

However, teleparallel gravity requires us to associate the teleparallel connection with each tetrad, which follows from our observation in Section~\ref{secmeaning} that, unlike in the general metric-affine case, in teleparallel geometries the tetrad and spin connection are not independent variables. We can then always find a local Lorentz transformation that transforms the connection to zero and hence effectively eliminate it from the theory and obtain the Weitzenb\"ock gauge.  The same local Lorentz transformation that eliminated the connection is now applied to the tetrad and transform it to the preferred class of tetrads.

There are different arguments in favor of the Weitzenb\"ock gauge as the fundamental definition of teleparallelism in the literature. One of them is  to follow the original Einstein's definition and postulate teleparallel geometry by the distant parallelism condition \eqref{Etelecond}.
The other  approach is  to start with the metric-affine viewpoint, where we do have both the tetrad and spin connection as independent variables, but then argue that since the teleparallel spin connection does not affect the dynamics of tetrads and does not have any field equations in the TEGR case, it can be set to zero \cite{Maluf:1994ji,Maluf:2013gaa}. We naturally agree with this possibility, but argue against viewing this as the  fundamental definition of teleparallel geometry, which was being asserted recently by Maluf \textit{et. al.} \cite{Maluf:2018coz}.

Another approach is what we will call the St\"uckleberg viewpoint, where the covariant formulation of teleparallel gravity is viewed as a  St\"uckelberg trick applied to the theory in the Weitzenb\"ock gauge 
\cite{BeltranJimenez:2017tkd,BeltranJimenez:2018vdo,BeltranJimenez:2019nns,BeltranJimenez:2019esp,Golovnev:2023yla}\footnote{We remind here that the St\"uckelberg trick was used originally in the massive electromagnetism and gravity, where the starting point is the gauge-invariant massless theory. Adding a mass term  breaks gauge invariance of the original theory, which is then restored by introducing the St\"uckelberg  fields by hand. In this picture,  the gauge symmetry is viewed as an artificial construction where any theory can be made gauge invariant by adding additional fields \cite{Hinterbichler:2011tt}. }
While often this viewpoint can be very useful to explore the dynamics of teleparallel theories \cite{BeltranJimenez:2019nns}, in the recent paper by Golovnev \cite{Golovnev:2023yla} it was argued that the theory in the Weitzenb\"ock gauge is the fundamental one   and the  local Lorentz symmetry  is ``artificially" restored by ``introducing" $\Lambda^a{}_b$  as St\"uckelberg  fields.

While the judgment of  whether something is artificial or not is ultimately subjective, we find it rather untenable to view $\Lambda^a{}_b$ as something artificially introduced on the manifold. Quite the contrary, the change of a basis is a fundamental concept connected with the very definition of the manifold, which is defined even before we introduce the metric structure. Teleparallel geometry  just takes this natural structure present on the manifold and defines the teleparallel connection as an unique connection given by the choice of the basis.
We find  the argument that the covariance is a fake or artificial symmetry  rather misleading since the whole point of the covariant derivative was to make-as the name suggests-the derivative covariant. The covariant derivative is indeed ``just" a St\"uckelbergization of a partial derivative, which, by the same logic, should be then viewed as an artificial concept.

Our argument here is that if we view the covariant formulation as just the artificial restoration of  symmetry using the St\"uckelberg trick, then the Weitzenb\"ock definition of the teleparallelism is the fundamental one and we should indeed  de-St\"uckelbergize our theory. 
However, then  there is no reason to call the coefficients of anholonomy the "teleparallel torsion tensor in the Weitzenb\"ock gauge". Or in the symmetric teleparallel case, where the  partial derivative of the metric is identified with the non-metricity tensor
\begin{equation}\label{key}
\Qsz_{\rho\mu\nu}=\partial_\rho g_{\mu\nu},
\end{equation}
to be called the ``symmetric teleparallel non-metricity tensor in the coincident gauge". 

While it is completely artificial to call the coefficients of anholonomy and partial derivatives of the metric  as tensors or view them  fundamentally as tensors in some gauge,
these are perfectly reasonable non-tensorial quantities, and we can construct theories of gravity in terms of them. We can either use them directly or equivalently consider their linear combinations, \eqref{lccon} and \eqref{Christoffel}, and work in terms of the Riemannian spin and linear connections. 

We choose the latter option and define the ``spin connection object"
\begin{equation}\label{key}
	\SCO=\ombol^{a}_{\,\,\,ca}\ombol^{bc}{}_{b}
	-\ombol^{a}{}_{cb}
	\ombol^{bc}{}_{a},
\end{equation}
and the  ``linear connection object" 
\begin{equation}\label{key}
	\LCO=\Gammabol^\rho{}_{\sigma\mu}\Gammabol^\sigma{}_{\rho\nu}-
	\Gammabol^\rho{}_{\mu\nu}\Gammabol^\sigma{}_{\rho\sigma},
\end{equation}
which are both quasi-scalars, i.e. their densities, $h\, \SCO$ and $\sqrt{-g}\, \LCO$, do transform by a total derivative term under the change of the non-coordinate and coordinate bases, respectively. It is then guaranteed that a Lagrangian given by a linear function of either of these connection objects will lead to the covariant field equations.

Of course, these are just the Einstein \eqref{LagEin} and M\o ller \eqref{LagMoll} Lagrangians, 
which are fully consistent gravitational Lagrangians. The quasi-invariance of these Lagrangians requires that they must be evaluated in some specific basis to avoid potentially divergent terms, and can be used to calculate the gravitational action that agrees with the standard GHY approach \cite{Krssak:2023nrw}.

The main claim of our paper is that  there are actually no covariant and non-covariant formulations of teleparallel theories. Only the covariant formulations are  the  genuinely teleparallel theories, where the geometry and gravity are described in terms of torsion or non-metricity, because only in this case we do actually  have the torsion and non-metricity tensors.
The reason why the ``non-covariant formulations" of teleparallel gravities work and produce  correct results in various situations is because the Einstein and M\o ller Lagrangians work. However, these are defined within the Riemannian geometry and do not use torsion or non-metricity at all. 

Our viewpoint is that teleparallel geometries are the  proper mathematical frameworks that naturally covariantize the Einstein and M\o ller Lagrangians. Taking away the covariance then defies their whole purpose and introduces serious  mathematical inconsistencies,  such as  misidentifying the coefficient of anholonomy with the torsion tensor. Of course, these mathematical inconsistencies can be resolved by simply acknowledging that we actually use the Riemannian connections and 	utilizing the Einstein and M\o ller Lagrangians.

We do believe that this viewpoint offers an interesting outlook on various modified gravity models. It should be straightforward to see that $f(\Tw)$ and $f(\Qs)$ gravity in the Weitzenb\"ock and coincident gauges  are  equivalent to   $f(\SCO)$ and $f(\mathcal{\LCO})$  gravity theories
\cite{Boehmer:2021aji,Blagojevic:2023fys} 
\begin{equation}\label{key}
\Lag_\SCO=\frac{h}{2\kappa} f(\SCO), 
\qquad
\Lag_\LCO=\frac{h}{2\kappa} f(\LCO).	
\end{equation}

We would like to argue here that this equivalence is trivial and we can actually extend it to  all teleparallel modified theories in the Weitzenb\"ock and coincident gauges, and view them as theories with Lagrangians given by various arbitrary contractions of the Riemannian connections with all indices summed over, which are not scalars or invariants in any meaningful sense. 

This brings us to the problem of modified teleparallel theories, all of which were constructed based on the claim that the 'torsion scalar' or other index-less objects used in their Lagrangians are some genuinely new teleparallel invariants not attainable within the standard Riemannian geometry \cite{Ferraro:2006jd,Ferraro:2011ks,Linder:2010py,Geng:2011aj,Bahamonde:2017wwk}. However, if we consider the gauge-fixed theories as fundamental ones, both of these claims are incorrect: these Lagrangians are not invariants or scalars, and it is possible to construct them within the Riemannian geometry using the connection coefficients.

We can then ask the question whether there  is any dynamical difference in the covariant approach, i.e. whether covariant  $f(\Tw)$ and $f(\Qs)$ differ from $f(\mathcal{\SCO})$ and  $f(\LCO)$ in any other way than ``just" being their covariant formulations. In the symmetric teleparallel case, the invariance under diffeomorphism symmetry gives us   extra 4 equations obtained by variation with respect to $\xi^\alpha$ \cite{BeltranJimenez:2018vdo}. However,  since these are the field equations for $\xi^\alpha$, they can be solved by choosing proper $\xi^\alpha$, i.e. these equations are the conditions on the coordinates and the coincident gauge. 
In the torsional case, it turns out that the field equations for $\Lambda^a{}_b$ are equivalent to the antisymmetric part of the field equations for the tetrad \cite{Golovnev:2017dox}. Therefore, we can observe that in both torsional and symmetric teleparallel gravity theories, the covariance gives us the field equations that determine a preferred basis. As soon as we determine this preferred basis, these theories are equivalent to their Riemannian connection analogues of the kind  of $f(\mathcal{\SCO})$ and  $f(\LCO)$ gravity.

The interesting question is whether the covariance could change the number of dynamical degrees of freedom. While in  the symmetric teleparallel theories as $f(\Qs)$ gravity, the number of degrees of freedom is not fully resolved issue yet \cite{Hu:2022anq,Tomonari:2023wcs,DAmbrosio:2023asf}, the topic is better understood in the case of torsional teleparallel theories. In  $f(\Tw)$ case, it is generally established that the number of propagating degrees of freedom is 5 \cite{Li:2011rn,Blagojevic:2020dyq}. Moreover, it was shown that the number of degrees of freedom should not be changed in the covariant formulation \cite{Ong:2013qja,Blixt:2022rpl}, which is analogous to the situation in New General Relativity (NGR) case \cite{Blixt:2018znp}. The NGR result can be easily understood using the  perturbative analysis, since the antisymmetric tetrad perturbations $u_{[\mu\nu]}$ and the perturbations of spin connection $w_{\mu\nu}$ appear in the field equations only in the combination $y_{\mu\nu}=u_{[\mu\nu]}-w_{\mu\nu}$, it is possible to made the whole analysis in terms of the symmetric perturbations of the tetrad and $y_{\mu\nu}$ \cite{Hohmann:2018jso}. Since $y_{\mu\nu}$ has the same form and field equations as $u_{[\mu\nu]}$, the number of degrees of freedom is then unchanged in the covariant picture. We would like to suggest here that the only possible way how the number of  degrees of freedom could change is if $y_{\mu\nu}=0$, i.e. if the spin connect degrees of freedom are chosen exactly to counter-balance the Lorentz degrees of freedom of the tetrad.

\section{Acknowledgements}
The author would like to thank Jose Geraldo Pereira for an invitation to submit this work and patience during the editorial process. This work was funded through  SASPRO2 project \textit{AGE of Gravity: Alternative Geometries of Gravity}, which has received funding from the European Union's Horizon 2020 research and innovation programme under the Marie Skłodowska-Curie grant agreement No. 945478.

\bibliography{references}
\bibliographystyle{Style}

\end{document}